\documentclass[twocolumn,aps]{revtex4}
\usepackage{amsmath}
\usepackage{amssymb}
\usepackage{graphicx}
\usepackage{dcolumn}
\usepackage{bm}
\usepackage{xcolor}
\usepackage{epstopdf}

\begin{document}

\preprint{Phys.Rev.B}

\title{Geometric engineering of viscous magnetotransport in a two-dimensional electron system}
\author{A. D. Levin,$^1$ G. M. Gusev,$^1$  A. S. Yaroshevich,$^{2,3}$ Z. D. Kvon,$^{2,3}$ and A. K. Bakarov $^{2,3}$}

\affiliation{$^1$Instituto de F\'{\i}sica da Universidade de S\~ao
Paulo, 135960-170, S\~ao Paulo, SP, Brazil}
\affiliation{$^2$Institute of Semiconductor Physics, Novosibirsk
630090, Russia}
\affiliation{$^3$Novosibirsk State University, Novosibirsk 630090,
Russia}

\date{\today}
\begin{abstract}
In this study, we present our experimental investigation on the magnetotransport properties of a two-dimensional electron system in GaAs quantum wells utilizing a variety of device geometries, including obstacles with thin barriers and periodic width variations. Our primary focus is to explore the impact of these geometries on the electron viscous flow parameters, enabling precise manipulation of hydrodynamic effects under controlled conditions. Through an analysis of the large negative magnetoresistivity and zero field resistivity, we deduce the scattering times for electron-electron and electron-phonon interactions, as well as the effective channel width. Our findings confirm that the system under investigation serves as a tunable experimental platform for investigating hydrodynamic transport regimes at temperatures above 10 K.

\end{abstract}

\maketitle
\section{INTRODUCTION}
The concept that has significantly enhanced our understanding of electronic transport phenomena is the notion that, when electron-electron scattering is strong enough, an effectively viscous hydrodynamics approach can be employed \cite{gurzhi}-\cite{andreev}. Gurzhi proposed this idea a while ago, but only recently have we been able to conduct a systematic investigation using a set of exceptionally clean samples that allow for the observation of a wide range of hydrodynamic effects. These effects include resistance decreasing  with  temperature (Gurzhi effect) \cite{gurzhi, dejong, andreev, principi}, giant negative magnetoresistance \cite{alekseev1, narozhny1, alekseev, narozhny2, gusev1, shi, raichev2, gusev5}, negative nonlocal resistance \cite{bandurin1, torre, pellegrino2, levin}, superballistic flow \cite{kumar, holder} and modifications to the Hall effect \cite{berdyugin, scaffidi, burmistrov, alekseev2, alekseev3, gusev2}. For a comprehensive overview of the field of viscous electronics, refer to papers \cite{polini}-\cite{narozhny}.

Viscous electron flows are expected to manifest in resistivity when the mean free path for electron-electron collisions (represented by $l_{ee}$) is considerably shorter than  the mean free path resulting from impurity and phonon scattering (denoted as $l$). Theoretical propositions suggest a direct proportionality between the electrical resistivity of a two-dimensional system and the electron shear viscosity, which can be expressed as $\eta=\frac{1}{4}v_{F}^2\tau_{ee}$, where $v_{F}$ represents the Fermi velocity, and $\tau_{ee}$ denotes the scattering time arising from electron-electron interactions, given by $\tau_{ee} = \frac{l_{ee}}{v_{F}}$.

Geometry plays an essential role in hydrodynamic flow. A Poiseuille geometry ($l_{ee} < W < l$)  allows for the establishment of a parabolic flow profile within the confined space.  In this scenario, hydrodinamic electron transport takes place, driven by the electric field, and encounters diffusive scattering at the channel's boundary. The relationship between resistivity and width is predicted to follow an inverse square law, where resistivity $\rho$ is inversely proportional to the square of the width ($\rho \sim W^{-2}$) \cite{gurzhi, du}. Similarly, resistivity is also expected to be inversely proportional to the square of the temperature ($\rho \sim T^{-2}$) \cite{gurzhi, gusev1, gusev4}. Importantly, a noticeable decrease in resistance as temperature rises has been observed in devices with an H-shaped geometry \cite{gusev1}. The idea that the Gurzhi effect could be connected to the unevenness in the velocity field due to the shape has been suggested. Exploring this phenomenon in devices with varying shapes would be valuable, potentially providing deeper insights into the electron hydrodynamic.
 The boundary conditions of the system can be described in terms of diffusive scattering or by introducing a slip length denoted as $l_{s}$. In extreme cases, the boundary conditions can be classified as "no-slip" ($l_s$ tends to zero) or "no-stress" ($l_{s}$ tends to infinity). When slip length approaches infinity (no-stress condition), it is anticipated that the Gurzhi effect will not be observed \cite{kiselev, raichev}.

An additional example is when a circular obstacle is present within the channel (Stokes geometry). Even if the slip length exceeds the size of the sample, there can still be the emergence of viscous shear forces, leading to the reappearance of the Gurzhi effect \cite{lucas, gusev3, krebs}. Furthermore, in a Stokes geometry, the pre-turbulent regime is predicted for large flow velocity as a periodic separation of hydrodynamic vortices, resulting in the formation of a phenomenon known as the Kármán vortex street \cite{goy}.
\begin{figure*}
  \centering
  \includegraphics[width=17cm]{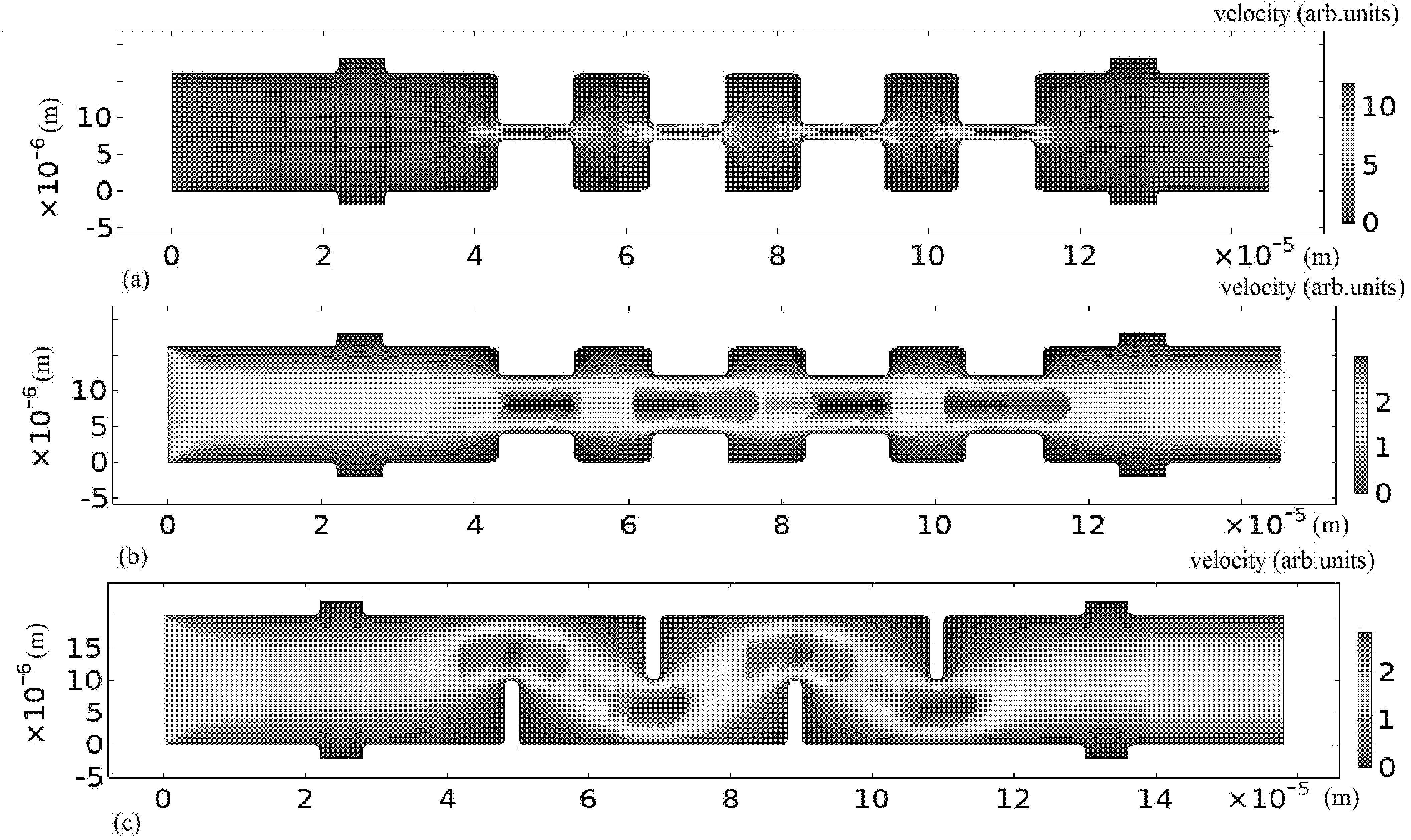}
\caption{(Color online) Hydrodynamic velocity flow. (a) Sketch of the velocity flow profile in a device with periodic periodic rectangular channel with narrow width of W=2 $\mu m$, configuration C1  (b) Sketch of the velocity flow profile in device with periodic rectangular channel width, with  narrow width of W=8 $\mu m$, configuration C2. (c) Sketch of the velocity ﬂow profile in
the presence of a set of thin barrier obstacles (zigzag barrier structure), configuration C3 The width of the sample is 20 $\mu m$.}
\end{figure*}

The  nonlinear effects have been theoretically explored within the confines of samples possessing Ventury geometry \cite{hui}. These peculiar samples are characterized by a continuous expansion of their width. Drawing a parallel to the hydrodynamic Bernoulli effect, it has been suggested  the potential utilization of hydrodynamic materials as a novel foundation for constructing nonlinear electronic devices \cite{hui,studenikin}.

The recent focus of the theoretical models' advancements has been on modifying slip parameters in a channel caused by a sequence of slender obstructions \cite{moessner}. Consequently, it becomes evident that not only the sample geometry itself, but also the geometry of its boundaries can exert an influence on transport properties, thereby enabling the advent of hydrodynamic conditions within narrow channels \cite{keser}.

Studying the magnetohydrodynamic behavior of electron transport significantly enhances our comprehension of viscous transport, enabling us to extract key parameters like electron-electron scattering rates and slip lengths \cite{alekseev1, alekseev, narozhny2, gusev1,  raichev2}. In simpler situations, the width of the sample directly factors into equations that describe magnetoresistance induced by viscosity. However, a more thorough analysis across a range of sample widths is essential for this theory. Additionally, the specific geometric arrangement may impact magnetoresistance. Our research is positioned to attract theoretical attention and could potentially serve as a foundational basis for future investigations.

In the current study, we have conducted experimental investigations on the transport properties of a mesoscopic 2D electron system in GaAs quantum wells with various geometries. Three distinct device configurations were examined ( see figure 1). The first configurations involve rectangular-like variations in the sample width, resulting in the formation of cavities after the electron flow traverses long, narrow constrictions (figure 1a,b), and are nominated as C1 and C2.  The other configuration (C3) consists of a series of obstacles with asymmetrically positioned barriers, enabling a zigzag-like flow pattern within the sample ( figure 1c).  For all configurations we observe a giant negative magnetoresistance at low magnetic field. By analysing  this pronounced negative magnetoresistivity and the resistivity in zero magnetic field, we  extract  the scattering times associated with electron-electron and electron-phonon interactions. Furthermore, we determine the effective width of the channel utilized in these experiments, which is found to be coincident with the geometric width within an order of magnitude variation.

\section{EXPERIMENTAL RESULTS}
We  used high-quality GaAs quantum wells for our devices. These wells have a width of 14 nm and an electron density of approximately $7.1 \times 10^{11} cm^{-2}$ at a temperature of 1.4 K. The mobility of the sample was $2\times 10^{6} cm^{2}/Vs$. To conduct our measurements, we designed a Hall bar specifically for multiterminal experiments. The sample consists of three consecutive segments with different lengths L (100, 20, and 100 $\mu m$), each being W= 20 $\mu m$ wide. Additionally, we incorporated eight voltage probes into the setup. Ohmic contacts to a two-dimensional electron system were fabricated by the annealing of the Ti/Ni/Au that is deposited on the GaAs surface. Ti/Au Schottky gates were fabricated to control electrostaticaly defined barriers in 2D liquid.  To create electrostatic barriers, we apply a gate voltage of $V_g = -0.9 V$.

For the measurements, we utilized a VTI cryostat and employed a conventional lock-in technique. This technique allowed us to measure  the longitudinal resistance. To avoid overheating effects, we applied an alternating current (ac) of 0.1-1 $\mu A$ through the sample, which is considered sufficiently low. The current I flows between contacts 1 and 4, and the voltage V was measured between probes 2 and 3, $R = R_{2,3}^ {1,4} = V_{2,3}/I_{1,4}$ (figure 2). Furthermore, we compared our findings with the transport properties of two-dimensional (2D) electrons in a larger-scale sample. The mean free path of electrons in macroscopic sample is $25 \mu m$ at T=4.2K, that exceeds the width of the sample.

\begin{figure}[ht]
\includegraphics[width=9cm]{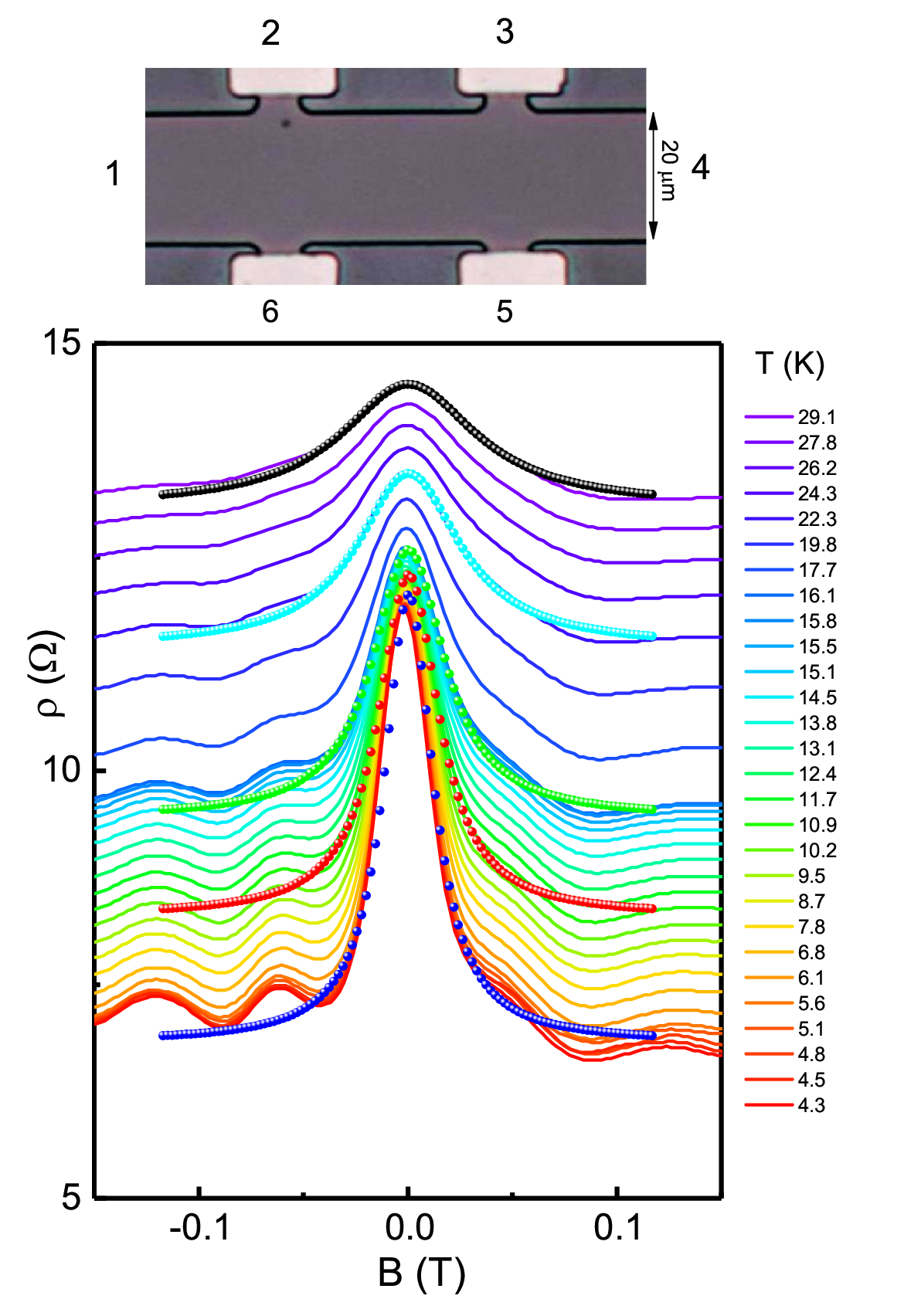}
\caption{(Color online)
Temperature-dependent  magnetoresistivity  of  unpatterned mesoscopic GaAs. The circles are examples illustrating magnetoresistance calculated from
 Eqs. (1) for different temperatures T(K): 4.3 (blue), 10.9 (red), 15.5 (green), 22.3 (cyan), 29.1 (black). Top-image of the central part of the Hall bar with 6 contacts.}
\end{figure}
In this paper, our main focus lies on conducting magnetoresistivity measurements and observing the resistivity behavior at zero magnetic field with varying temperature, particularly for different geometries. We begin by conducting measurements on unpatterned samples. Figure 2 illustrates the resistivity ($\rho=\frac{W}{L}R$) evolution as a function of magnetic field at different temperatures and the image of the device. It is notable that there is a significant negative magnetoresistivity ($\rho(B)-\rho(0) < 0$) characterized by a Lorenztian profile, which becomes smaller and broader as the temperature increases. Additionally, the resistivity at zero magnetic field exhibits an increase with temperature. This observation agrees with previous findings, which were interpreted as distinctive characteristics of hydrodynamic behavior \cite{alekseev1, gusev1, gusev2, gusev4} except for the temperature range $4.2 < T < 10$ K. In this temperature interval, both ballistic and hydrodynamic properties should be considered equally in describing the system's behavior \cite{raichev2}. The small oscillations observed above $B>0.05 T$ for low temperatures in Figure 2. The Larmor radius is 3.5 $\mu m$ at $B=\pm
0.06 T$ and $1.6 \mu m$ at $B=\pm0.12 T$, both of which correspond to the maxima of the oscillations. These values are notably smaller when compared to the dimensions of the sample in terms of width or length. The observed oscillations can potentially be attributed to a degree of commensurability with the width of the potentiometric probe, as well as the misalignment of the gold contacts with the sample edges. It's important to note that these oscillations are minimal and do not significantly impact the magnetoresistance, particularly at higher temperatures.
\begin{figure}[ht]
\includegraphics[width=9cm]{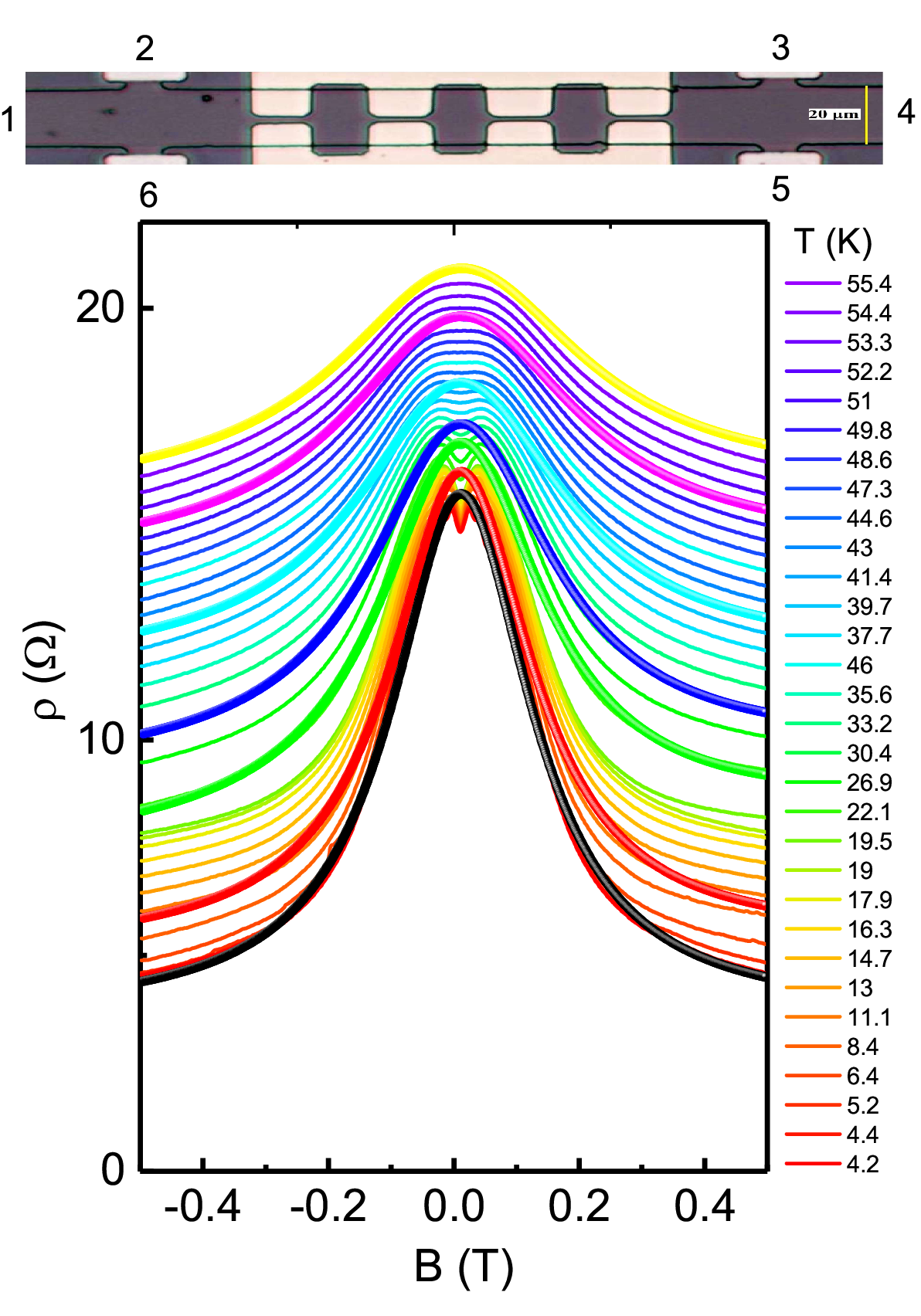}
\caption{(Color online)
Temperature-dependent magnetoresistivity of  a mesoscopic GaAs  for configuration C1. The circles (thick lines) are examples illustrating magnetoresistivity calculated from
 Eqs. (1) for different temperatures T(K): 4.2 (black), 11.1 (red), 22.1 green) , 30.4 (blue), 41.4 (cyan) , 51 (magenta), 55.4 (yellow). Top-image of the central part of the Hall bar with 6 contacts.}
\end{figure}

\begin{figure}[ht]
\includegraphics[width=9cm]{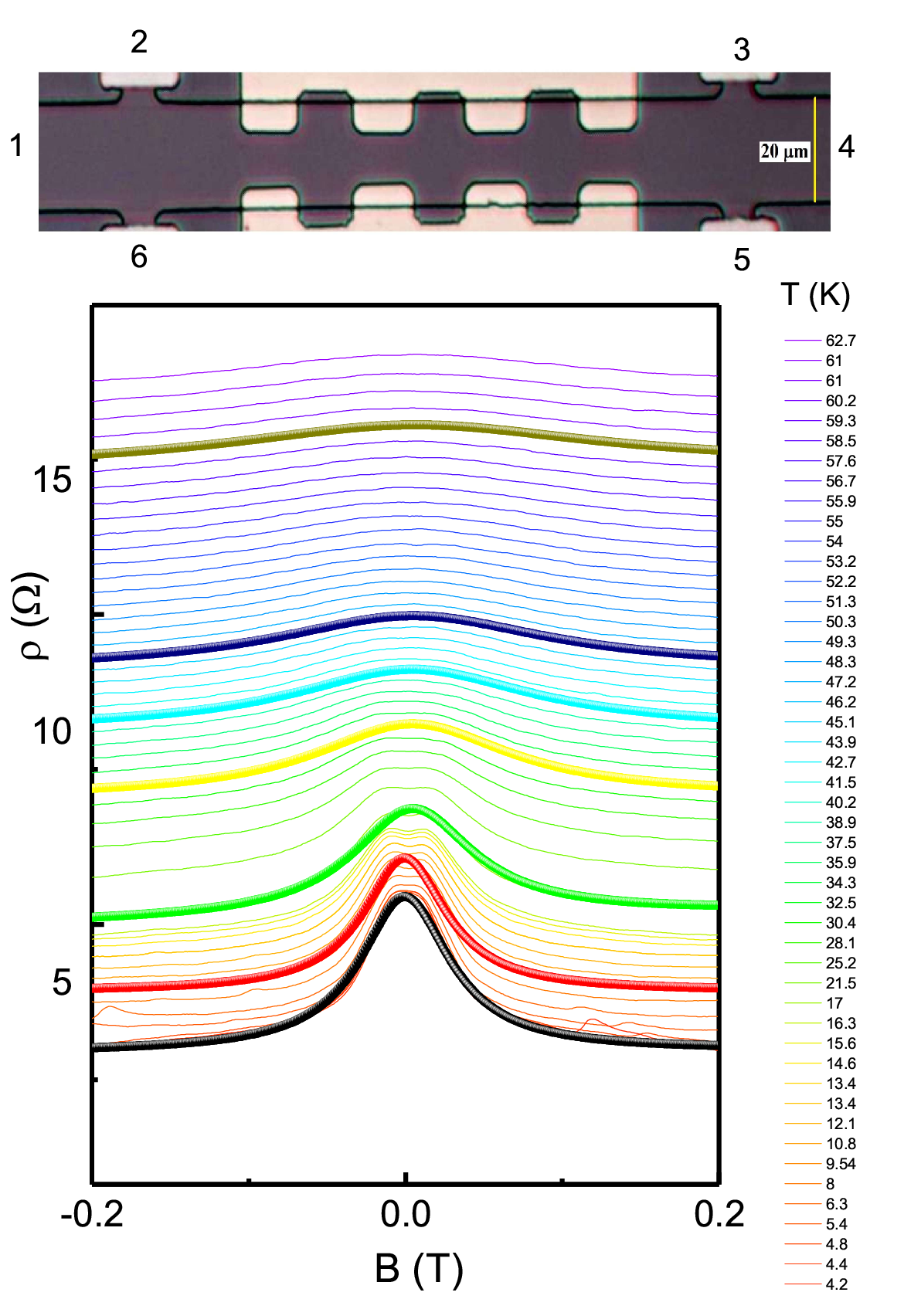}
\caption{(Color online)
Temperature-dependent  magnetoresistivity  of  a mesoscopic GaAs  for configuration C2. The circles (thick lines) are examples illustrating magnetoresistivity calculated from
 Eqs. (1) for different temperatures T(K): 4.2 (black), 9.5 (red), 17 (green), 32.5 (yellow), 40.2 cyan), 46.2 (violet), 59.3 (olive). Top-image of the central part of the Hall bar with 6 contacts.}
\end{figure}

\begin{figure}[ht]
\includegraphics[width=9cm]{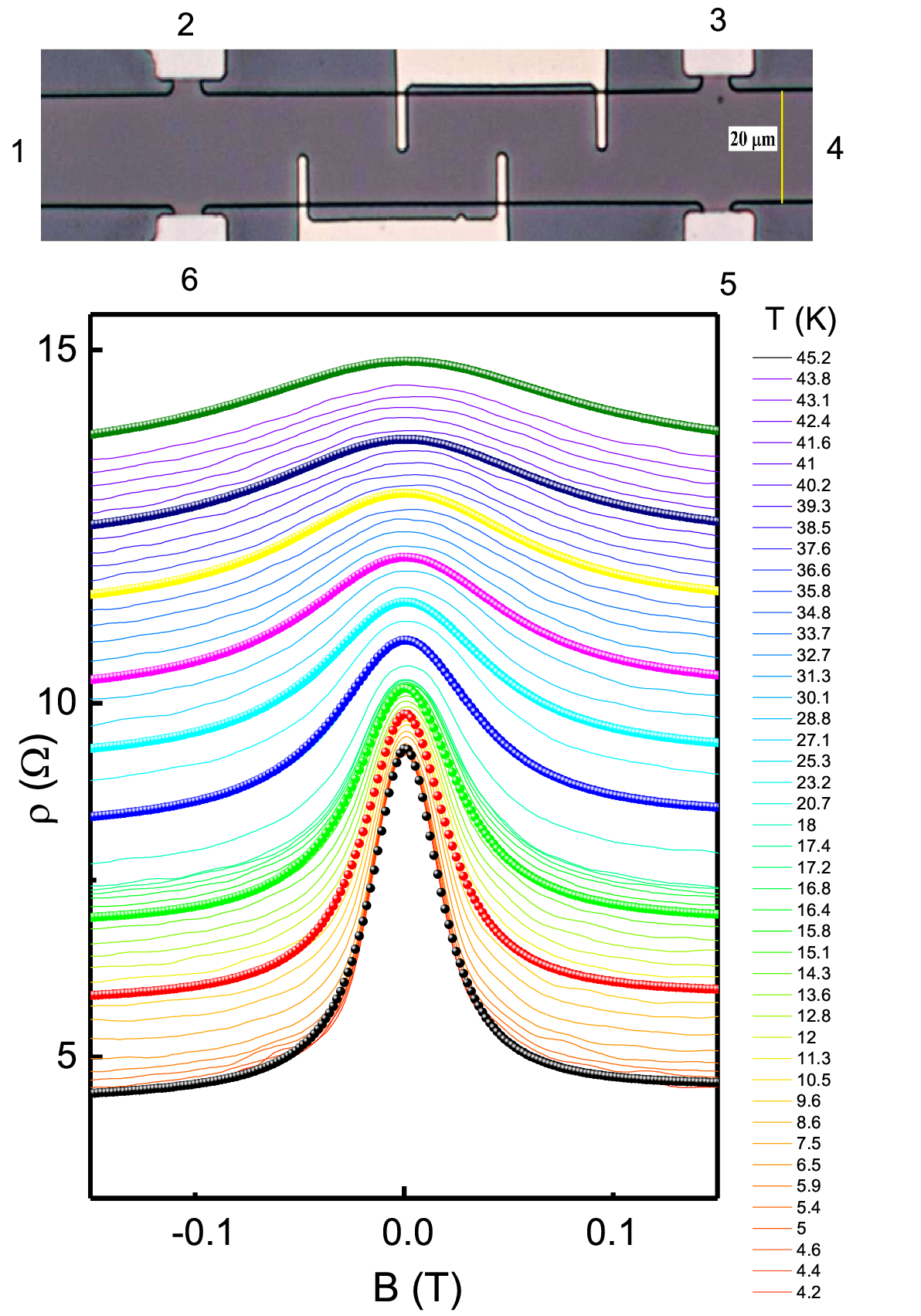}
\caption{(Color online)
Temperature-dependent  magnetoresistivity  of  a mesoscopic GaAs  for configuration C3. The circles (thick lines) are examples illustrating magnetoresistivity calculated from
 Eqs. (1) for different temperatures T(K): 4.2 (black), 10.5 (red), 15.1 (green), 20.7 (blue), 25.3 (cyan)), 30.1 (magenta), 35.8 (yellow), 40.2 (violet), 45.2 (slimy green). Top-image of the central part of the Hall bar with 6 contacts.}
\end{figure}
\begin{figure*}
  \centering
  \includegraphics[width=15cm]{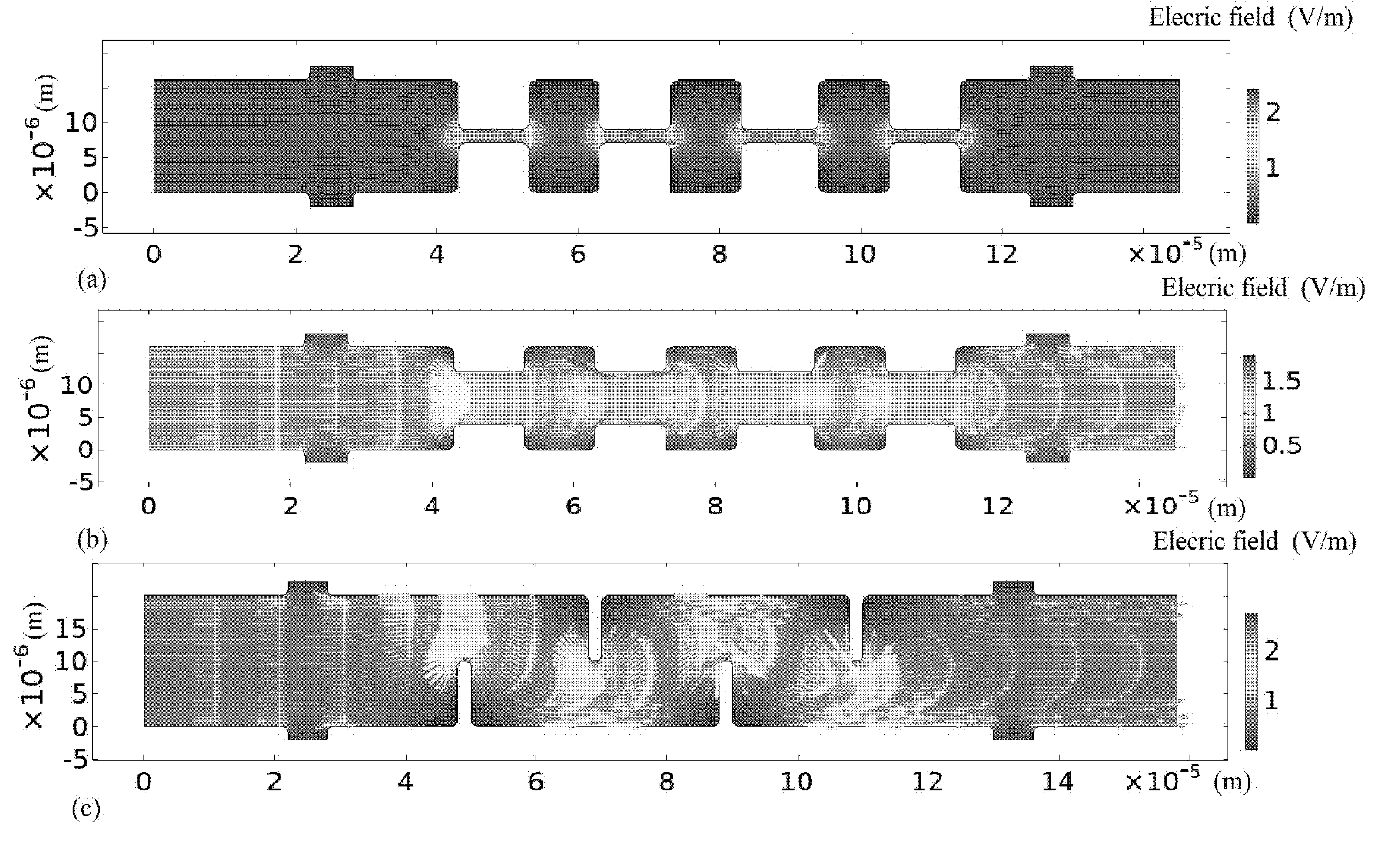}
\caption{(Color online) Ohmic current flow. (a) Sketch of the electric field profile in a device with periodic periodic rectangular channel width narrow width of W=2 $\mu m$, configuration C1  (b) Sketch of the electric field profile in a device with periodic rectangular channel width, with  narrow width of W=8 $\mu m$, configuration C2. (c) Sketch of the electric field profile in
the presence of a set of thin barrier obstacles (zigzag barrier structure), configuration C3 The width of the sample is 20 $\mu m$.}
\end{figure*}

Figures 3, 4, and 5 illustrate the pronounced negative magnetoresistivity observed in configurations C1, C2, and C3, respectively. The figures also show the image of the devices, demonstrating configuration of the barriers. It is important to note that both the width and height of the Lorentzian profile exhibit strong dependence on the specific configurations. For instance, the magnetoresistivity for C1 configuration appears significantly broader compared to C2 and C3 geometries. Furthermore, the Lorentzian profile for C2 geometry exhibits the smallest height among the three configurations. As the temperature increases, the peaks become broader while maintaining a Lorentzian profile across all devices. However, it is important to note that for all configurations, the peak at zero magnetic field consistently increases with temperature. This observation indicates the absence of the Gurzhi effect and suggests that disorder and  phonon scattering contributes more significantly to the resistivity compared to hydrodynamic effects for these configurations. In order to provide a more comprehensive understanding of this behavior, we conduct a detailed comparison with theoretical models in the next section.
\section{THEORY AND DISCUSSION}
In current theories, electron transport in mesoscopic samples is typically analyzed using various models such as ballistic, hydrodynamic, or more general frameworks (see for review \cite{narozhny}). These models are based on a detailed approach that involves solving the Boltzmann kinetic equation while considering boundary conditions for the electron distribution function.

In our study, we employ the model proposed in Refs. \cite{alekseev1, scaffidi} as it encompasses the essential magnetohydrodynamic properties, including the intricate effects associated with the relaxation of the distribution function's second harmonic by defects and electron-electron scattering.

This model presents conductivity as a combination of two independent contributions. The first contribution is attributed to ballistic effects or static disorder, while the second contribution arises from viscosity \cite{alekseev1}. The approach involves utilizing a magnetic field-dependent viscosity tensor and deriving the resistivity tensor:

\begin{equation}
\rho(B)= \frac{m}{e^{2}n}\left(\frac{1}{\tau}+\frac{1}{\tau^{*}}\frac{1}{1+(2\omega_{c}\tau_{2})^{2}}\right),\,\,\,
\end{equation}

where $1/\tau$ is the scattering  rate due to static disorder, $m$ and $n$ are the effective mass and the density, where $\omega_{c}=\frac{eB}{mc}$ is the cyclotron frequency, $\tau^{*}=\frac{W^{2}}{12\eta}$, where $\eta=\frac{1}{4}v_{F}^{2}\tau_{2}$ is the viscosity. The shear viscosity relaxation rate is given by
\begin{equation}
\frac{1}{\tau_{2}(T)}=\frac{1}{\tau_{2,ee}(T)}+\frac{1}{\tau_{2,imp}}=A_{ee}\frac{(kT)^{2}}{\hbar E_{F}}+\frac{1}{\tau_{2,imp}},
\end{equation}

where $A_{ee}$ is numerical factor which can be different for the weakly and strongly interacting Fermi system \cite{alekseev4}. The relaxation rate $\frac{1}{\tau_{2,imp}(T)}$, which arises from any process responsible for relaxing the second harmonic of the distribution function, including scattering by static defects, contributes to viscosity. On the other hand, $\frac{1}{\tau_{2,ee}(T)}$ corresponds to the relaxation of shear viscosity due to electron-electron scattering \cite{alekseev, alekseev1}. The momentum relaxation rate is given by

\begin{equation}
\frac{1}{\tau(T)}=\frac{1}{\tau_{0, ph}(T)}+\frac{1}{\tau_{0, imp}},
\end{equation}

where $\tau_{0, ph}$ represents the term associated with phonon scattering, and $\tau_{0, imp}$ represents the scattering time resulting from static disorder (not related to the relaxation time of the second moment) \cite{alekseev1, alekseev}.

In our experiments, we typically measure the resistance directly, and the resistivity is subsequently calculated using geometric factors. Specifically, the resistivity $\rho$ is determined using the equation $\rho=\frac{W}{L}R$, where R is the resistance, W represents the width and L denotes the length of the rectangular device. For configurations C1, C2, and C3, the simple relation mentioned earlier is not applicable due to the presence of barriers. In order to calculate the resistivity, we solve the Laplace equation for potentials while specifying appropriate boundary conditions. This allows us to determine the effective ratio $W_{eff}/L_{eff}$, which is used to calculate the resistivity in these cases. In the Ohmic case, Figure 6 illustrates the profile of the electric field for configurations C1, C2, and C3. Indeed the equipotential lines consistently (not shown) intersect the borders at right angles, while the current flows in the direction of decreasing potential. Furthermore, unlike hydrodynamic flow ( figure 1), the current density profile across the sample width does not exhibit a parabolic shape.

Based on the calculated potential distribution, we determine the resistivity. Subsequently, we perform a fitting of the magnetoresistance curves and the $\rho (T)$ at zero magnetic field shown in Figures 2-5. The fitting is done using three parameters: $\tau (T), \tau ^{*} (T)$ and $\tau_{2} (T)$. Excellent agreement with equation 1 is observed over a wide range of magnetic fields and temperatures. Furthermore, we demonstrate that the two parameters,$\tau ^{*} (T)$ and $\tau_{2} (T)$, are not completely independent but instead maintain a constant ratio with respect to the sample width.
\begin{figure}[ht]
\includegraphics[width=9cm]{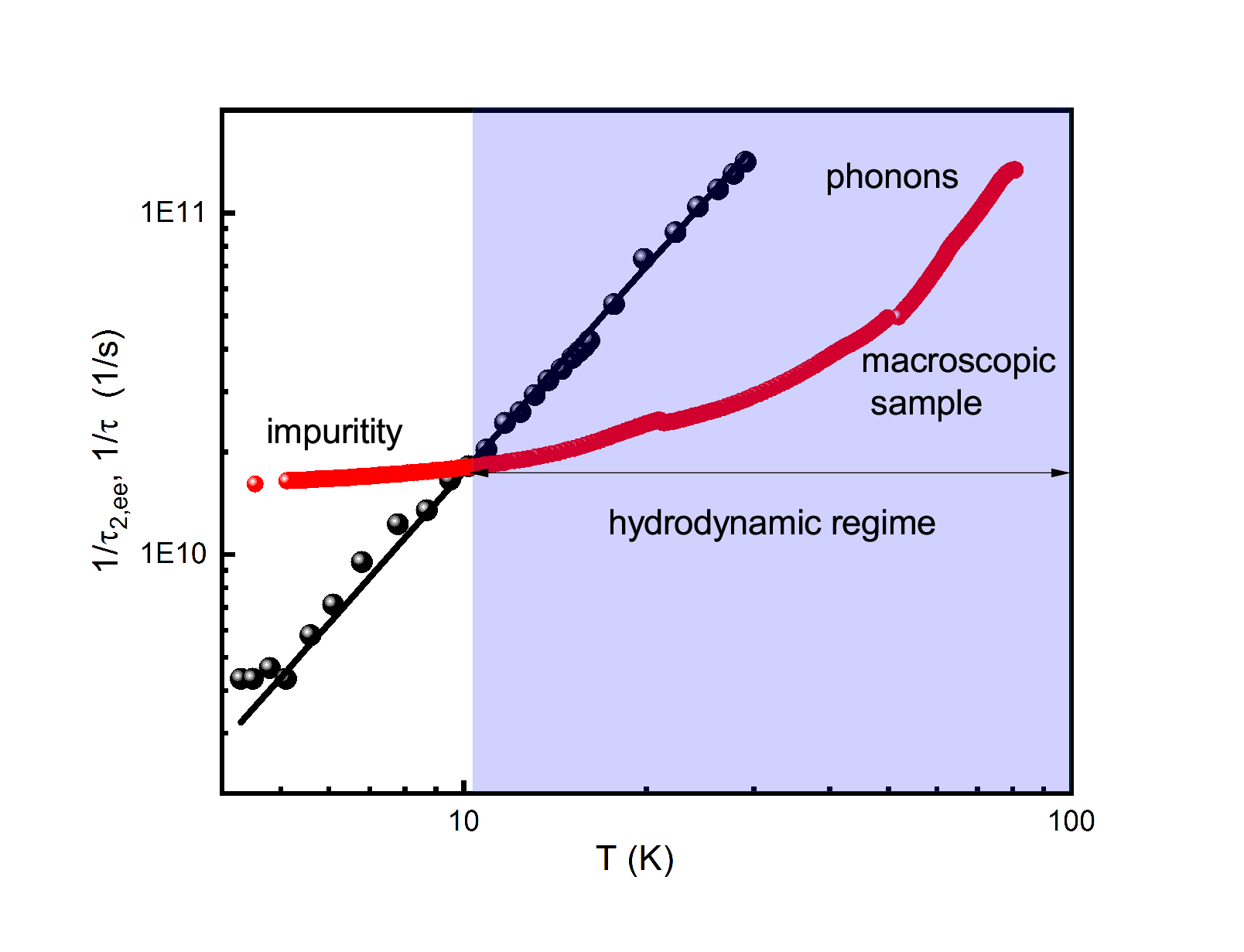}
\caption{(Color online)
The relaxation rate, represented by black circles, denoted as $1/\tau_{2,ee}$, is obtained by comparing with experimental data in an unpatterned sample. The impurity scattering rate, represented by red circles, denoted as $1/\tau$, is derived from the macroscopic sample mobility and is plotted as a function of temperature. The blue shading highlights the temperature range where $1/\tau_{2,ee} > 1/\tau$, indicating the presence of the hydrodynamic regime. At low temperatures, scattering is predominantly governed by static impurities, whereas at higher temperatures, scattering by phonons becomes more significant. }
\end{figure}
Now, let's shift our focus to the data concerning electron-electron interaction, which can be derived from the analysis of magnetoresistance. Figure 7 displays the data for $1/\tau_{2,ee}$, which is determined through the comparison of the magnetoresistivity curves with equation 1 in an unpatterned sample. This parameter is associated with inelastic electron-electron scattering, as indicated in equation 2. In addition, we include the dependence of $1/\tau (T)$ extracted from the macroscopic sample mobility for comparison. As mentioned in the introduction, the hydrodynamic regime is expected to be relevant when the electron-electron collision rate is significantly higher than the scattering rate due to impurities and phonons. This specific region is highlighted with a blue shading indicating temperature interval above $T>10 K$. We can see here that  $1/\tau_{2,ee}$ follows $T^{2}$ behaviour in accordance with equation 2. Parameters $A_{ee}$  extracted from this comparison are indicated in Table 1. By comparing the temperature dependency of the relaxation rate $1/\tau_{2,ee}$  with equation (2), we can deduce a temperature-independent characteristic time $1/\tau_{0,imp}$ (table 1). The hydrodynamic approach is linked to a significant relaxation of the mth harmonic of the distribution function caused by disorder scattering with the rates $1/\tau_{m,imp}$ \cite{alekseev1, narozhny}.

\begin{figure}[ht]
\includegraphics[width=9cm]{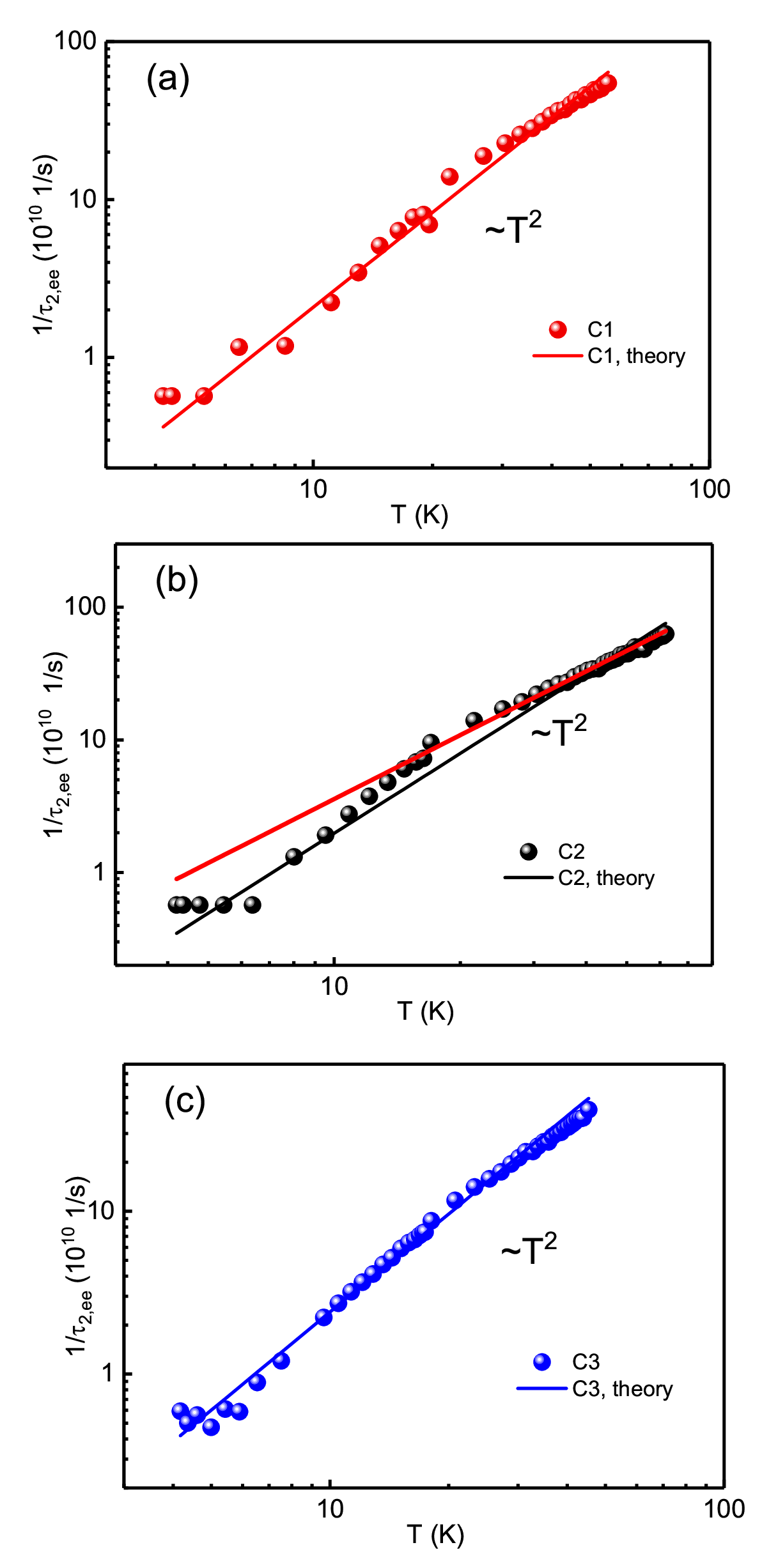}
\caption{(Color online)
The relaxation rate,  $1/\tau_{2,ee}$ a a function of the temperature obtained for different configurations: (a)-C1, (b)-C2, (c)-C3. Solid lines- theory. }
\end{figure}
\begin{figure}[ht]
\includegraphics[width=9cm]{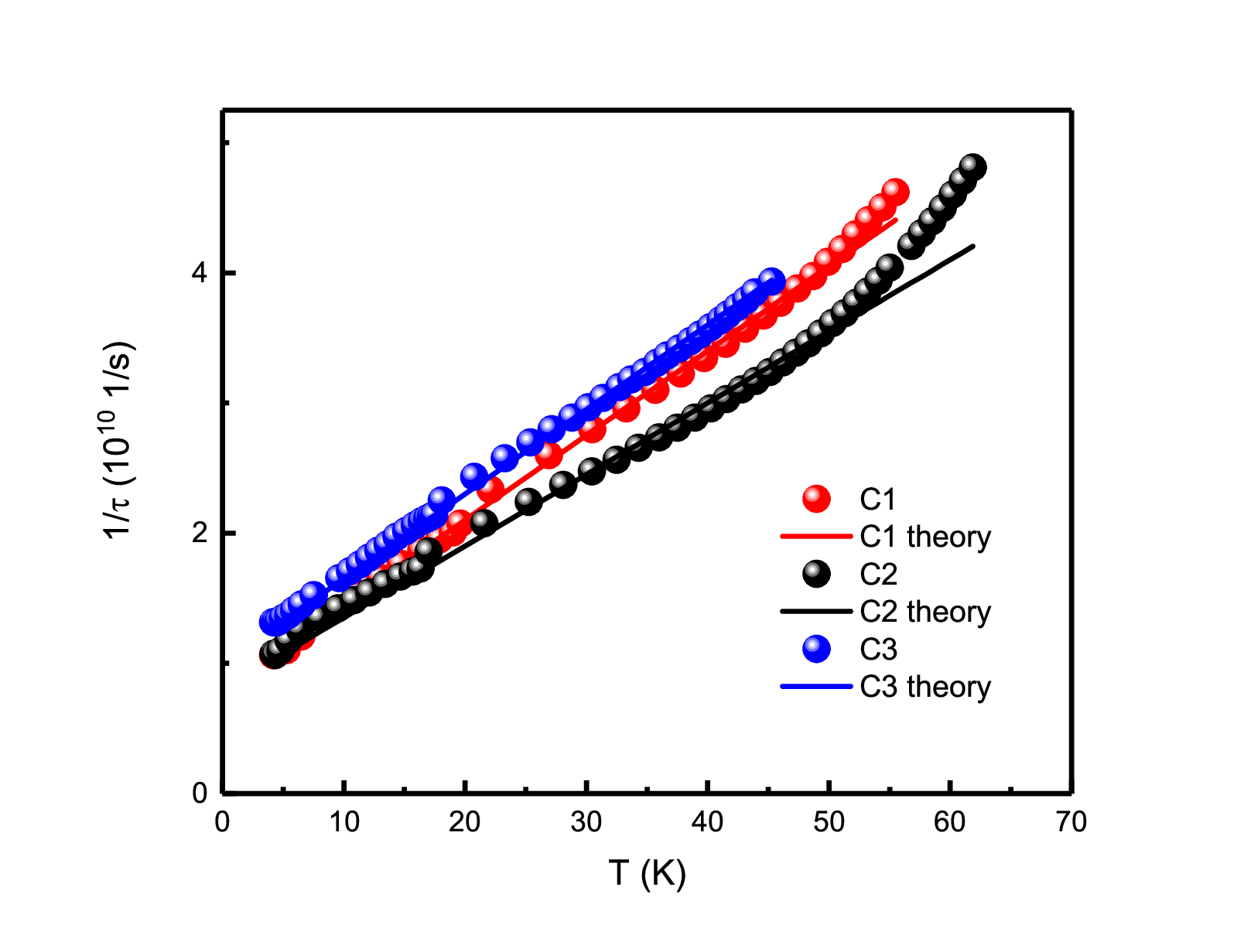}
\caption{(Color online)
The momentum relaxation rate,  $1/\tau$, as a function of the temperature obtained for different configurations. Solid lines- theory.}
\end{figure}
\begin{figure}[ht]
\includegraphics[width=9cm]{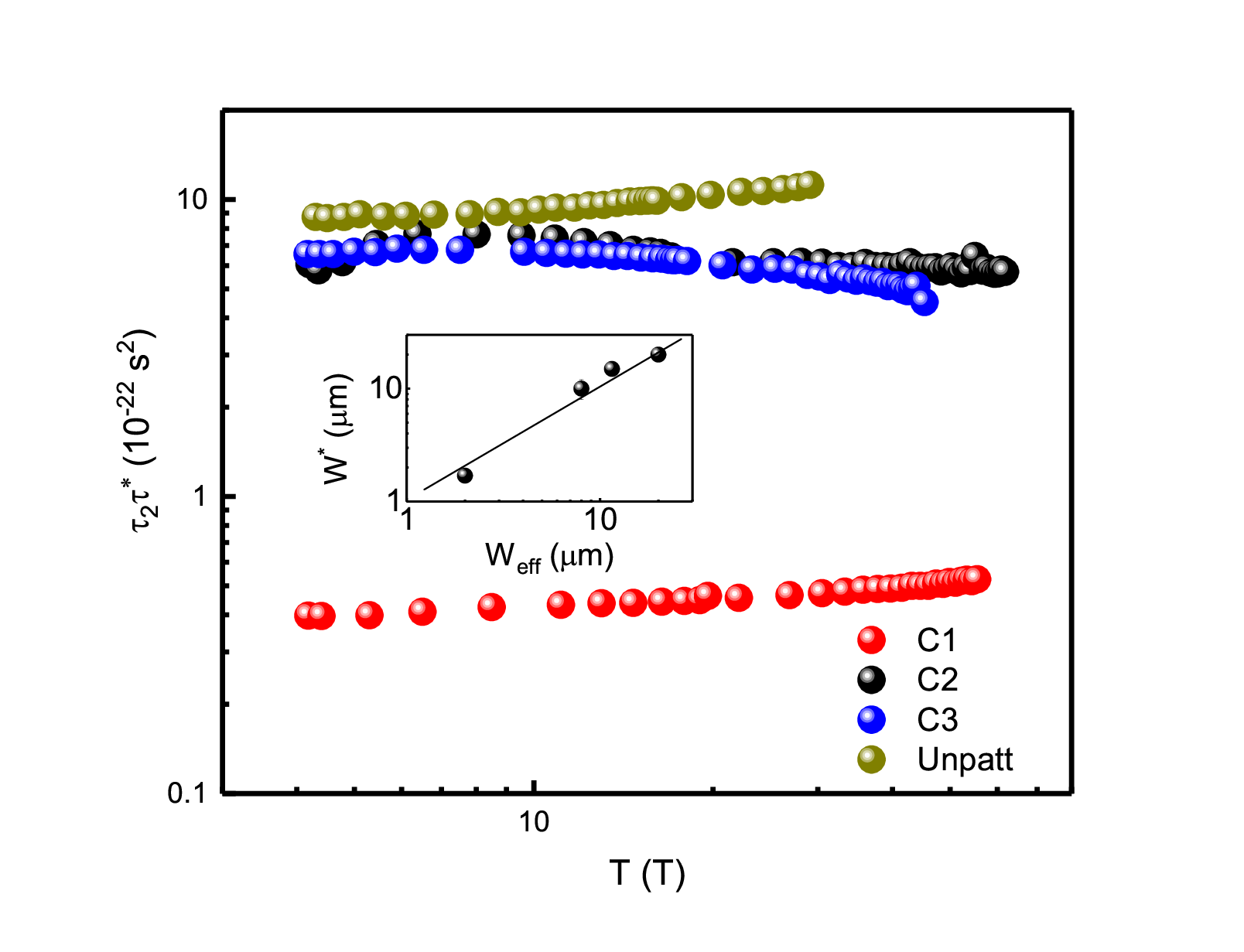}
\caption{(Color online)
 The product of the relaxation times $\tau_{2,ee}\tau^{*}$ as a function of the temperature.
 This product is proportional to $W^{2}$ and allows to extract the width of the channel $W^{*}$.
 The insert shows the correspondence between effective geometrical width $W_{eff}$ and   $W^{*}$}.
\end{figure}
In Figures 8 and 9, a overview of the extracted parameters is presented, including the relaxation rate $1/\tau_{2,ee}$ (a) and the momentum relaxation rate $1/\tau$ (b) plotted against temperature. To facilitate a comparison with the theoretical predictions, we employed equations 2 and 3 and expressed the rates as $1/\tau_{2,ee}=A_{ee}\frac{(kT)^{2}}{\hbar E_{F}}$ and $1/\tau=B_{ph}T+1/\tau_{0, imp}$. Parameters $A_{ee}, B_{ph}, \tau_{0, imp}, \tau_{2, imp}$ are presented in the table 1.
\begin{table}[ht]
\caption{\label{tab1} Fitting parameters of the electron system  for different configurations. Parameters are defined in the text.}
\begin{ruledtabular}
\begin{tabular}{lccccccc}
&Config. &$1/\tau_{2,imp}$&$1/\tau_{0,imp}$ & $A_{ee}$ & $B_{ph}$ & $W_{eff}$&$W^{*}$   \\
& & $(10^{11} 1/s)$ & $(10^{10} 1/s)$ &   & ($10^{9} 1/sK$) & $\mu m$  & $\mu m$  \\
\hline
&Unpatt. & $0.7$  & $1.7$ & $0.53$ &  $0.7$ & 20& 20\\
&C1& $6.95$  & $0.8$ & $0.6$ &  $0.65$ & 1.7& 1.5\\
&C2& $1.5$  & $0.8$ & $0.6$ &  $0.55$ & 6& 10\\
&C3& $1.0$  & $1.0$ & $0.73$ &  $0.65$& 11.5 & 15 \\
\end{tabular}
\end{ruledtabular}
\end{table}
Despite the seemingly distinct behavior of magnetoresistivity depicted in figures 2-5, characterized by varying Lorentzian heights, widths, and different values of $\rho(B=0)$, it is observed that all relaxation rates converge onto universal curves: $1/\tau_{2,ee}\sim T^{2}$ and $1/\tau \sim T$, exhibiting nearly identical parameters. The parameters $A_{ee}$ and $B_{ph}$ represent the rates of electron-electron and electron-phonon scattering, respectively. These parameters align with previously extracted values and correspond to the theoretical models \cite{alekseev1}.
Upon analyzing these dependencies in figure 8, we hold the perspective that, contrary to an exponential decline for temperatures exceeding 30K in configurations C1 and C2, there seems to be a small bump in these trends ( less significant for C3) . We contend that this observation could be linked to challenges in fitting the data rather than a shift in the relaxation mechanism. Our approach involved applying a simplified theory of magnetoresistance for a rectangular sample, whereas our actual configuration is more intricate.

It is worth mentioning that the effective time $\tau^{*}$, obtained from the height of the Lorentzian profile in the magnetoresistivity, is inversely proportional to the relaxation time $\tau_{2}$ that determines the width of the Lorentzian profile. Consequently, the product of these two quantities is anticipated to be temperature-independent and can be expressed as follows:
\begin{equation}
\tau_{2}\tau^{*}=\frac{W^{2}}{3v_{F}^{2}}
\end{equation}
This equation enables us to independently determine the effective channel width. Figure 10 shows the product $\tau_{2,ee}\tau^{*}$ as a function of temperature. Notably, this parameter exhibits minimal temperature dependence over a wide temperature range. Specifically, in the case of the unpatterned sample and configuration C1, a slight increase is observed, while configurations C2 and C3 demonstrate a decrease in this parameter. Indeed, the derived channel width, denoted as $W^{*}$, closely aligns with the geometrical width over a wide range of variation spanning approximately one order of magnitude. Note, that we determine from Ohmic current distribution (fig.6) geometrical width $ W_{eff}$ for C3 configuration (zigzag-like). This agreement unequivocally justifies that our magnetoresistivity arises from viscosity and has a hydrodynamic origin. One would expect that the dependence on temperature would be temperature-independent in the case of ballistic or classically-sized magnetoresistivity, at least until the mean free path exceeds the width of the sample. Furthermore, it is highly likely that the temperature dependence would differ for different widths for $l=v_{F}\tau > W$. However, what we observed is a universal temperature dependence characterized by a $T^{-2}$ broadening of the Lorentzian shape, which is more indicative of electron-electron scattering rather than the $T^{-1}$ dependence associated with momentum relaxation due to phonon scattering.

\begin{figure}[ht]
\includegraphics[width=9cm]{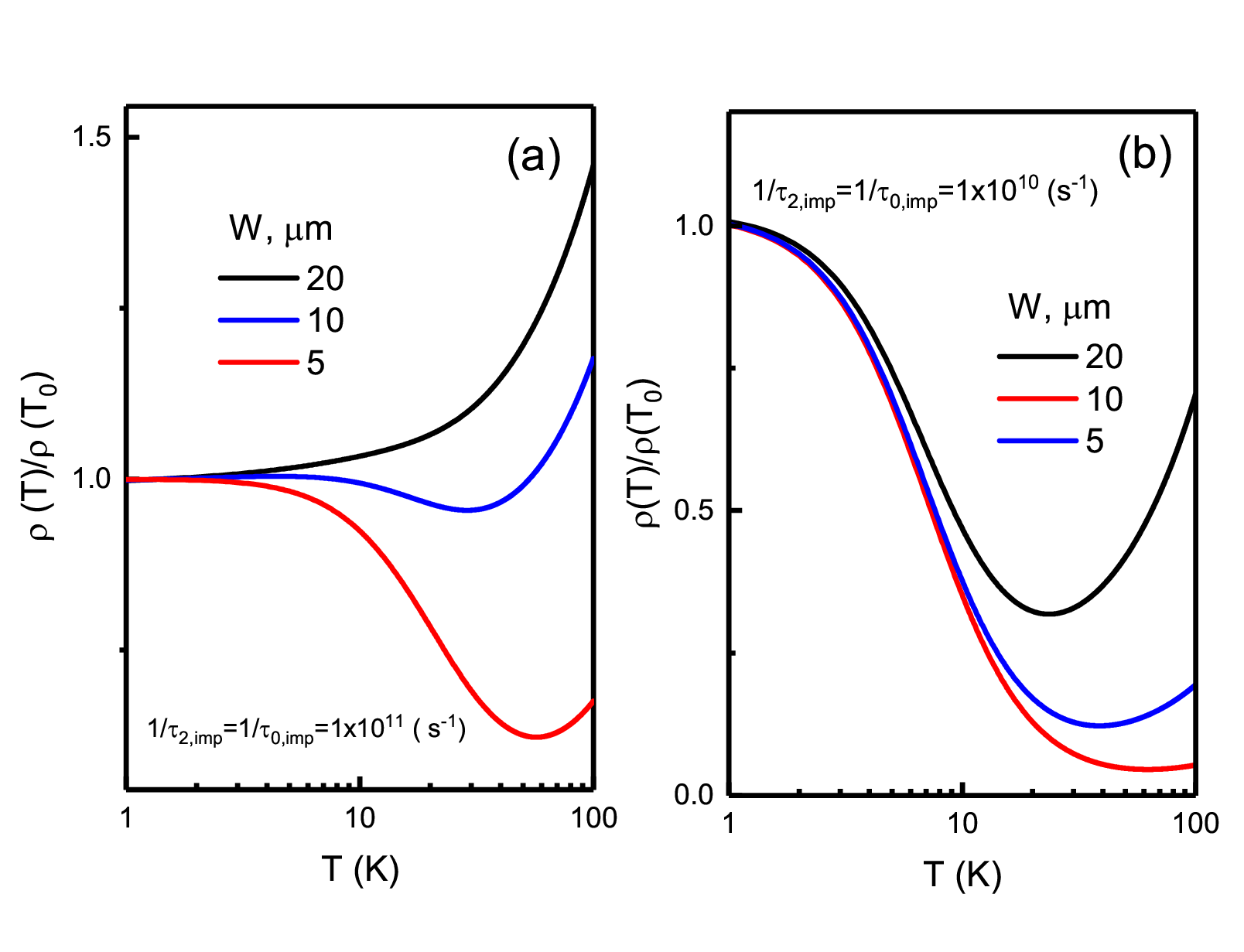}
\caption{(Color online)
The relative resistivity in zero magnetic field, as determined by equation 1 using the parameters $A_{ee}=0.6$ and $B_{ph}=0.6\times10^9 \frac{1}{\text{sK}}$, is plotted as a function of temperature for various sample widths, $T_{0}=1K$. The corresponding relaxation times are indicated on the panel for figs. (a) and (b).}
\end{figure}

In the concluding section of the paper, we direct our attention towards an important question that was initially investigated by Gurzhi. We explore which parameters, associated with realistic samples, can lead to a more notable phenomenon: a decrease in resistivity with increasing temperature \cite{gurzhi}. By examining Table 1, it becomes apparent that different configurations and sample geometries allow us to alter the conditions for the hydrodynamic effect, resulting in a substantial variation of the negative magnetoresistance. Interestingly, despite these variations, fundamental parameters associated with electron-electron collisions and scattering by phonons, such as $A_{ee}$ and $B_{ph}$, exhibit universality and remain consistent. It is important to note that despite the significant variations in parameters, we did not observe the Gurzhi effect in our devices. Instead, the resistivity in all devices exhibited an increase with temperature. We calculate the resistivity in zero magnetic field by using equation 1 for universal parameters  $A_{ee}=0.6$ and $B_{ph}=0.6\times10^9 \frac{1}{\text{sK}}$ varying the width of the sample and relaxation time of the second  harmonic of
distribution function caused by disordered scattering $\tau_{2,imp}$. The results of these calculations are depicted in Figures 11a and 11b. Interestingly, we observe a significant dependence of the Gurzhi effect on a particular parameter. As illustrated in the figures, only when the relaxation rate $1/\tau_{2,imp}$ is relatively small, specifically below $0.7\times 10^{11} s^{-1}$, do we observe a decrease in resistivity with temperature, particularly in narrower devices. For larger relaxation rates, only an increase in resistivity is expected. This explains why we did not observe the Gurzhi effect in the samples examined in this study. According to the parameters listed in the table, the relaxation rate $1/\tau_{2,imp}$ is relatively large, which prevents the observation of the Gurzhi effect. Unfortunately, even in narrower samples, the relaxation rate $1/\tau_{2,imp}$ is significantly increased, which unfortunately suppresses the Gurzhi effect. In our previous study a pronounced Gurzhi effect has been observed due to a combination of small  $1/\tau_{2,imp}$ and narrow width \cite{gusev1, gusev4}. The importance of the relaxation rate $1/\tau_{0,imp}$ is relatively low, but it is desirable to have a value smaller than $10^{11} s^{-1}$ for the observation of the Gurzhi effect. Exploring the microscopic nature of the relaxation rate $1/\tau_{2,imp}$ would be intriguing as it could enhance the hydrodynamic conditions for manipulating the Gurzhi effect. The potential existence of a hydrodynamic regime in real samples seems to be connected to the significant relaxation of both odd and even harmonics in electron scattering on disorder, with relaxation rates denoted as $1/\tau_{m,imp}$ that are much larger for odd harmonics with $m \geq 3$ compared to $1/\tau_{m,ee}$ for even harmonics. If relaxation only occurred due to electron-electron scattering, a substantial difference in relaxation times between even and odd harmonics could result in the emergence of anomalous non-hydrodynamic transport regimes \cite{gurzhi}.  Both contributions $1/\tau_{2,ee}$ and $1/\tau_{2,imp}$ to the relaxation rate $1/\tau_{2}$ are proportional to the products of the Landau parameter factor $(1 + F_{2})$ and the quasiparticle collision integrals averaged by energy \cite{alekseev}. In summary, although the Gurzhi effect is a delicate phenomenon that necessitates specific parameter combinations in real samples, the hydrodynamically induced negative magnetoresistance is highly robust to these parameters. It can be observed across a variety of geometric configurations, making it an invaluable tool for investigating hydrodynamic effects in various materials, including graphene and Dirac fermions in HgTe samples  \cite{berdyugin, gusev5}.

\section{CONCLUSION}

In this study, we conducted experimental investigations on the magnetotransport properties of a two-dimensional electron system in GaAs quantum well using different device geometries. We observed that the resistivity at zero magnetic field consistently increased with temperature, although the temperature dependence, represented by $\rho(T)$, varied for different configurations. We proposed that the Gurzhi effect, characterized by a decrease in resistivity with temperature increase, is governed by the relaxation rate of the second harmonics of the distribution function due to disorder.

On the other hand, we found that the hydrodynamically induced large negative magnetoresistivity was persistent across all geometries. By analyzing this pronounced negative magnetoresistivity and the resistivity in the absence of a magnetic field, we were able to extract the scattering times associated with electron-electron and electron-phonon interactions. Furthermore, we determined the effective width of the channel used in these experiments, which closely matched the geometric width with only a modest variation within an order of magnitude.

\section{ACKNOWLEDGMENTS}
 The financial support of this work by  FAPESP (Brazil),  CNPq (Brazil) and Ministry of Science and Higher Education of the Russian Federation is acknowledged.

\end{document}